\documentstyle[twoside,fleqn,npb,epsfig]{article}

\newcommand{\AmS}{{\protect\the\textfont2
  A\kern-.1667em\lower.5ex\hbox{M}\kern-.125emS}}

\title{NLO corrections to the polarized Drell-Yan cross section
       in proton-proton collisions}

\author{W.L. van Neerven\address{Instituut-Lorentz, University of Leiden
        P.O. Box 9506, 2300 RA Leiden, The Netherlands}%
        \thanks{Work supported by the EC network 'QCD and Particle 
         Structure' under contract No. FMRX-CT98-0194.}}

\begin{document}

\begin{abstract}
We present the full next-to-leading order (NLO) corrected inclusive
cross section for massive lepton pair production in longitudinally polarized
proton-proton collisions. All QCD partonic subprocesses have been included 
provided the lepton pair is created by a virtual photon, which is a valid
approximation for $Q<50~{\rm GeV}$. Like in unpolarized proton-proton
scattering the dominant subprocess is given by the $q(\bar q)g$-channel
so that massive lepton pair production provides us with an excellent
method to measure the spin density of the gluon. Using our calculations
we give predictions for the longitudinal spin asymmetry measurements at the
RHIC.
\end{abstract}

\maketitle

\section{Introduction}

At this moment the NLO calculations of unpolarized quantities are almost
finished so that one now is concentrating on the computations
of the next-to-next-to-leading (NNLO) corrections. In the case of polarized
processes this stage is not reached yet and in this contribution
we will report on a recent calculation of the complete NLO contribution 
\cite{Rav02} to massive lepton pair 
production (Drell-Yan process). From the study of the
Regge pole model in the sixties we have learnt that the predictions
are in better agreement with experiment when the reaction only
involves unpolarized particles. In the case the particles become polarized,
predictions and data are very often at variance with each other. The same also
seems to happen for perturbative QCD and therefore it will be very interesting
to study polarized reactions, as will be measured in the future at
the RHIC (BNL, USA), because they can provide us with a deeper insight in QCD.
At this moment there are only data available from polarized deep inelastic
lepton hadron scattering. However like in unpolarized scattering they
provide little information about the gluon and sea-quark densities which
are important quantities under study. In our contribution we will mainly
focus on the former density which can be much better extracted if
the hadronic reaction is dominated by partonic subprocesses with a gluon
in the initial state.
Processes which are suitable to extract the polarized gluon density are
jet production, charm quark production in photon-hadron collisions
or deep inelastic lepton-hadron scattering and direct photon production.
In the subsequent part of this paper we will concentrate under which
conditions the polarized gluon density can be extracted from 
semi-inclusive massive lepton pair production in proton-proton collisions.
The latter process is given by
\begin{eqnarray}
 p+p \rightarrow l^+ +l^- + 'X'\,,
\end{eqnarray}
where $X$ denotes an inclusive hadronic final state.
The reaction above is described by the polarized cross section 
\begin{eqnarray}
&& \frac{d^3\Delta \sigma^{pp}}{dQdp_Tdy}=
\nonumber\\[2ex]
&& \sum_{a,b=q,g}\Delta f_a^p(\mu^2)
\otimes \Delta f_b^p(\mu^2)\otimes \frac{d^3\Delta \sigma_{ab}(\mu^2)}{dQdp_Tdy}
\,.
\nonumber\\
\end{eqnarray}
Here $Q$ denotes the invariant mass of the lepton pair which has the
transverse momentum $p_T$ and rapidity $y$. Further $\Delta \sigma_{ab}$
and $\Delta f_a(\mu^2)$ represent the polarized partonic cross section 
and polarized parton density respectively which both depend on the
factorization scale $\mu^2$.
Notice when the value of $Q$ is
sufficiently small, e.g. $Q\ll M_Z$, the reaction in Eq. (1)
is dominated by a virtual photon $\gamma^*$ in the intermediate state
so that one can neglect the $Z$ contribution. 

In lowest order (LO) of the strong coupling constant
$\alpha_s$ the following partonic subprocesses contribute
to the cross section in Eq. 2
\begin{eqnarray}
&& q + \bar q \rightarrow g + \gamma^*\,,
\\[2ex]
&& g + q(\bar q)\rightarrow q(\bar q) + \gamma^*\,.
\end{eqnarray}
In next-to-leading order (NLO) one has to compute the one-loop 
contributions to the Born reactions appearing in the equation above and
the two to three parton subprocesses
\begin{eqnarray}
&& q + \bar q \rightarrow g + g + \gamma^*\,,
\\[2ex]
&& g + q(\bar q)\rightarrow g + q(\bar q) + \gamma^*\,,
\\[2ex]
&& q_1 + q_2\rightarrow q_1 + q_2 + \gamma^*\,,
\\[2ex]
&& q_1 + \bar q_2\rightarrow q_1 + \bar q_2 + \gamma^*\,,
\\[2ex]
&& g + g\rightarrow g + g + \gamma^*\,,
\end{eqnarray}
where the (anti-)quarks in Eqs. (7), (8) can be identical $q_1=q_2$
or non-identical $q_1\not =q_2$. Notice that the one-loop corrections
to the Born reaction in Eq. (3), the subprocess in Eq. (5) and the interference
term appearing in the $qq$-channel in Eq. (7) were calculated in
\cite{Chan98} which are in agreement with our results in \cite{Rav02}.
In \cite{Rav02} we have included the remaining contributions so that
at this moment the complete NLO correction to the cross section in Eq. (2)
is known. The outline of this calculation and the results predicted
for the RHIC experiments will be presented in the next section.
\section{Regularization in ${\bf n}$ dimensions with the ${\bf \gamma_5-}$
matrix}

The computation of the virtual contributions and the radiative corrections
to the partonic cross sections reveals the presence of ultraviolet,
infrared and collinear divergences in loop and phase space integrals. 
The usual method to regularize these singularities is given by 
$n$ dimensional regularization. The advantage of this method is that
before one has to carry out renormalization and mass factorization
all Ward-identities are automatically preserved. However this is not longer 
true when the $\gamma_5$-matrix and the Levi-Civita tensor appear.
The latter quantities show up in electro-weak interactions and in
the computation of polarized processes. The terms which violate the
Ward identities are called evanescent and they have to be removed
before renormalization and mass factorization are carried out.
One of the most chosen prescription for the $\gamma_5$-matrix is given by
the HVBM approach \cite{hvbm}. This prescription violates the Ward identity
for the non-singlet axial vector current and the Adler-Bardeen theorem
\cite{Adl69} so that one needs evanescent counter terms. The HVBM method
is rather complicated since it requires that the $n$-dimensional space
has to split up in a 4 and an $n-4$ dimensional subspace. Accordingly
the gamma-matrices and the momenta have to be split up which complicates
the gamma-matrix algebra and the phase space integrals. This will complicate
the calculations in particular if one wants to compute NNLO corrections.
To avoid this complication we have chosen the approach in \cite{Ak73}
and replace the $\gamma_5$-matrix by
\begin{eqnarray}
\gamma_{\mu}\,\gamma_5&=&\frac{i}{6}\,\epsilon_{\mu\rho\sigma\tau}\,
\gamma^{\rho}\, \gamma^{\sigma}\,\gamma^{\tau}\,,\quad \mbox {or} 
\nonumber\\[2ex]
\gamma_5&=&\frac{i}{24}\,\epsilon_{\rho\sigma\tau\kappa}\,\gamma^{\rho}\,
\gamma^{\sigma}\,\gamma^{\tau}\,\gamma^{\kappa} \,.
\end{eqnarray}
In this way one can apply the usual gamma-matrix algebra in $n$ dimensions.
Moreover the integration over the final state momenta is the same as
in processes where the $\gamma_5$-matrix and the Levi-Civita tensor
do not appear. Furthermore we contract the Levi-Civita tensors in
four dimensions before the phase space integrals are carried out.
We checked that in this procedure the matrix elements are independent
of an arbitrary axial gauge vector $l$ which appears in the polarization sum
\begin{eqnarray}
&& \sum_{\alpha=L,R} \epsilon^{\mu}(p,\alpha)\,
\epsilon^{\nu}(p,\alpha)=-g^{\mu\nu}+\frac{l^{\mu}\,p^{\nu}}
{l\cdot p}
\nonumber\\[2ex]
&&+\frac{l^{\nu}\,p^{\mu}}{l\cdot p}\,, \quad \mbox{with}\quad l^2=0\,.
\end{eqnarray}
This method leads to more evanescent counter terms than shown by the
HVBM approach. They can be extracted from a more simple cross section
than the one given in Eq. (2). Notice that the ultraviolet divergences
do not need evanescent counter terms because in process (4) only the
virtual photon is attached to the loop graphs. Therefore the coupling 
constant renormalization can be performed in the usual 
${\overline {\rm MS}}$-scheme and no additional evanescent counter term
is needed. Only the collinear divergences
which are removed by mass factorization
\begin{eqnarray}
&& d\,\Delta \hat \sigma_{ij}\left (\frac{1}{\varepsilon}\right )
=\sum_{k,l=q,g}\Delta \Gamma_{ki}
\left (\frac{1}{\varepsilon},\mu^2\right )
\nonumber\\[2ex]
&& \otimes
\Delta \Gamma_{lj}\left (\frac{1}{\varepsilon},\mu^2\right )
\otimes \Delta \sigma_{kl}(\mu^2)\,,
\end{eqnarray}
with
\begin{eqnarray}
&&\Delta \Gamma_{ij}=
\nonumber\\[2ex]
&&\delta_{ij}+\frac{\alpha_s}{2\pi}\left [\left (
\frac{2}{\varepsilon}+\gamma_E-\ln 4\pi\right )\,\Delta P_{ij}\right ]\,,
\end{eqnarray}
need evanescent counter terms for all four splitting functions $\Delta P_{ij}$.
The evanescent counter terms for the splitting functions $\Delta P_{qq}$
and $\Delta P_{qg}$ are extracted from the Drell-Yan polarized 
cross section $d\Delta \sigma/dQ$ of the processes in Eqs. (3) and (4)
respectively. This is achieved by comparing the coefficient functions
using our method above with the ones
obtained from a four dimensional regularization scheme where there is no 
problem with the $\gamma_5$-matrix and the Levi-Civita tensor. For instance
one can regularize the collinear divergences 
by taking the external quark and gluon legs off-shell ($p^2<0$) or the quark
gets a mass $m$ and one puts the external legs on-shell. 
In this case the kernels are given by
\begin{eqnarray}
\Delta \Gamma_{ij}=A_{ij}(p^2,m^2,\mu^2)=\langle j(p) |O_i|j(p)\rangle\,,
\end{eqnarray}
where $A_{ij}$ ($i,j=q,g$) denote the renormalized operator matrix elements
corresponding to the local operators appearing in the operator product 
expansion for the product of two electromagnetic currents. To obtain
the evanescent counter terms for $\Delta P_{gq}$ and $\Delta P_{gg}$
we followed the same procedure for the total cross section for polarized
Higgs production given by the subprocesses
\begin{eqnarray}
q + g \rightarrow q + H\,, \qquad g + g\rightarrow g + H\,.
\end{eqnarray}
The genuine ${\overline {\rm MS}}$-scheme for mass factorization is
now given by the following replacement in Eq. (13)
\begin{eqnarray}
\Delta P_{ij} \rightarrow \Delta P_{ij} + \quad \mbox{evanescent counter term}
\,,
\end{eqnarray}
where $\Delta P_{ij}$ can e.g. be found in \cite{Alt77}. In order to check
that the same evanescent counter terms also apply to the cross section
in Eq. (2) we recalculated the latter using a four dimensional regularization
method. It turned out that the evanescent counter terms are the same
for both the reactions in Eqs. (5)-(9) and the processes mentioned above.
\section{Results}
The hadronic cross section in Eq. (2) has been plotted using the following
input. For the C.M. energy of proton-proton collisions at the RHIC
we have chosen $\sqrt{S}=200~{\rm GeV}$. Further we adopted the NLO 
approximation for the running coupling constant and the polarized parton 
densities given by the parametrizations in \cite{Glu01}, \cite{Blu02}.
For the factorization scale, which is set to be equal to the renormalization
scale, we have taken $\mu^2=p_T^2+Q^2$.
Both parametrizations are presented in two scenarios depending on the
size of the gluon density. The parametrizations in \cite{Glu01} are
represented by the valence scenario (VS) and the standard scenario (SS).
Those in \cite{Blu02} are given by scenarios 1 (S1) and 2 (S2) respectively.
Over the whole $x$-region the polarized gluon densities for all four
scenarios approximately satisfy the following inequalities
\begin{eqnarray}
\Delta f_g^{VS}(x) < \Delta f_g^{SS}(x) <\Delta f_g^{S2}(x) <
\Delta f_g^{S1}(x) \,.
\end{eqnarray} 
From this behaviour it turns out that the $qg$-channel dominates the cross
section\\ $d^2\Delta \sigma^{pp}/dQ/dp_T$ provided $p_T>Q/2$. An exception 
is the VS-scenario where the $q\bar q$ subprocess becomes of equal
importance over the whole $p_T$-range. This feature was already discovered
in LO in \cite{Berg98} so that the NLO corrections do not change this picture.
However both scenarios (VS) and (SS) lead to the same transverse momentum
distributions so that one cannot distinguish them. The same holds for the
S2 scenario in \cite{Blu02} of which the cross section is slightly larger
than the ones given by VS and SS. Only the S1 scenario leads to a larger
cross section than the other ones. This is revealed by Figure 1 where
we have plotted in NLO the longitudinal asymmetry defined by
\begin{eqnarray}
A_{LL}=\frac{d^2\Delta \sigma^{pp}/dQ/dp_T}{d^2\sigma^{pp}/dQ/dp_T} \,,
\end{eqnarray}
for $Q=6~{\rm GeV}$. For the computation of the unpolarized cross
section in the denominator we have chosen the GRV98 set in \cite{Glu98}
with the same factorization scale as given above. From this figure we
infer that one cannot distinguish between the VS and SS scenario.
At $p_T=20~{\rm GeV/c}$ the difference between both scenarios and the
S2 scenario can be observed when the polarized cross section is
known up to 12.5$\%$ assuming 100$\%$ polarization for the proton beams
which is very unlikely. However when we allow for a 25$\%$ uncertainty
in the polarized cross section one can distinguish between scenarios S1 and S2
even if the protons are not fully polarized (e.g. 75$\%$). In Figure 2
we have studied the effect of the NLO corrections to the longitudinal
asymmetry when compared with the LO approximation. The effect is rather small
for the SS and S1 scenarios but amounts to 30-40 $\%$ at large $p_T$
for the VS and S2 scenarios. From this one can conclude that the
K-factors for the polarized and unpolarized cross sections are about the
same in particular for scenarios SS and S1.
\begin{figure}[htb]
\vspace{4.9cm}
\includegraphics{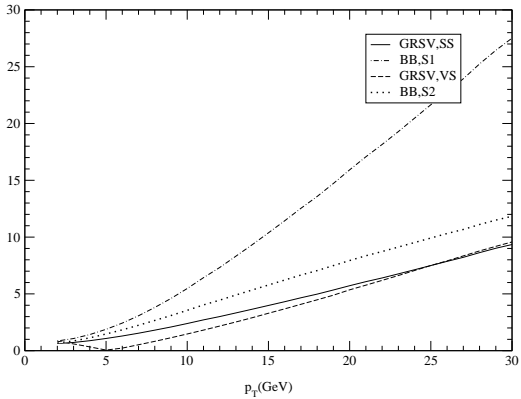}
\caption{Longitudinal asymmetry $A_{LL}$ in percentage.}
\label{fig:fig1}
\end{figure}
\begin{figure}[htb]
\vspace{5.5cm}
\includegraphics{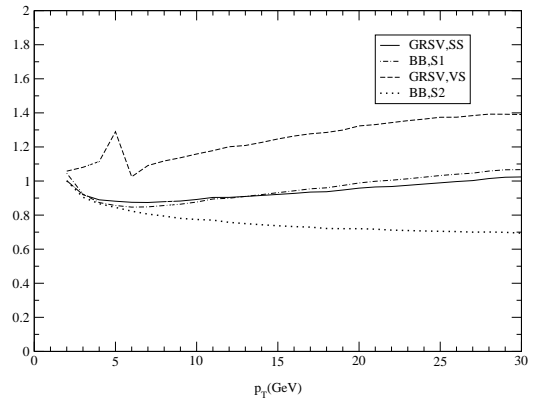}
\caption{Ratio $A_{LL}^{NLO}/A_{LL}^{LO}$.}
\label{fig:fig2}
\end{figure}

\end{document}